# The BOS-Lig Dataset: Accurate Ligand Charges from a Consensus Approach for 66,810 Experimentally Synthesized Ligands

Roland G. St. Michel[1,2,], Ryan J. Jang[2], Aaron G. Garrison[3], Ilia Kevlishvili[3], and Heather J. Kulik[1,3,*]

[1]*Department of Materials Science and Engineering, Massachusetts Institute of Technology, Cambridge, MA 02139, USA*

[2]*Department of Chemistry, Massachusetts Institute of Technology, Cambridge, MA 02139, USA*

[3]*Department of Chemical Engineering, Massachusetts Institute of Technology, Cambridge, MA 02139, USA*

*corresponding author email: hjkulik@mit.edu

ABSTRACT: Understanding ligand properties is essential for computational high-throughput screening of transition metal complexes. However, ligand properties such as net charge and other information such as their application area are often absent or inconsistently recorded in crystallographic datasets. Here, we construct a ligand dataset from 126,985 mononuclear transition metal complexes curated from the Cambridge Structural Database. Using an iterative charge-balancing workflow that combines complex charges, metal oxidation states, and consensus across crystallographic observations, we confidently assign net charges to 66,810 ligands among 94,581 identified unique ligand structures to curate the Boston Open-Shell Ligand (BOS-Lig) dataset. The workflow assigns ligand charges in homoleptic complexes first and then iteratively propagates these assignments across heteroleptic environments, allowing charges to be inferred even when direct charge information is unavailable. We analyze cases where simple heuristics such as the octet rule would have failed and introduce a purity metric to identify when our charge assignments may be incorrect. Each ligand is also classified in terms of its metal coordinating atoms and whether there are multiple variants (i.e., hemilability). We then link complexes to their associated journal abstracts and apply a topic-modeling workflow to link 25,146 ligands with functional application areas spanning reactivity, redox chemistry, biological chemistry, and photophysical chemistry. Together, we provide an experimentally grounded dataset of ligand chemical space that connects charge and functional application as a foundation for computational screening and data-driven ligand design.

## 1. Introduction

Transition metal complexes (TMCs) are widely studied for applications spanning catalysis[1-4], photochemistry[5-8], magnetism[5,9-12], and bioinorganic chemistry[13-15]. Their utility arises from the wide range of metal identities, accessible oxidation states, and spin configurations, combined with the chemical diversity of coordinating ligands[14,16,17]. Ligands play a central role in determining complex geometry, electronic structure, and reactivity, and even small changes in ligand identity or coordination can lead to substantial differences in function[18-21]. Because a given complex can vary simultaneously in metal identity, coordination geometry, and ligand composition, the number of distinct metal–ligand combinations grows rapidly[22-24]. As a result, the space of experimentally realized TMCs reflects a combinatorially large and chemically rich ligand landscape.

The increasing availability of crystallographic data has created new opportunities to study this chemical space systematically. Large experimental databases, such as the Cambridge Structural Database (CSD),[25] now contain hundreds of thousands of characterized TMCs and capture decades of synthetic effort across disparate chemical fields. However, extracting structures and properties from this data to enable computational screening remains challenging. Fundamental properties such as net charge and coordination behavior, as well as other information such as their functional role or application area, are often not explicitly encoded and must instead be inferred from incomplete or inconsistent metadata[26-28]. These inferences are frequently made using simplified heuristics that work well for narrow classes of complexes but generalize poorly across the full diversity of transition metal chemistry.

One persistent challenge is the assignment of individual ligand charges. While prior work has introduced scalable strategies for inferring ligand charges from crystallographic data, these approaches are often limited to specific coordination geometries or rely on assumptions that restrict their chemical coverage[29-33]. In recent years, additional tools have been developed to tackle this challenge. For example, cell2mol[34] interprets crystallographic unit cells by reconstructing molecular connectivity and charge assignments from atomic coordinates, but it may fail to process structures with large or unconventional unit cells. Reliable ligand charge assignment requires broadly applicable methods that reconcile conflicting observations across different depositions while remaining robust to experimental noise/error and structural variability. Robust ligand charge assignments are particularly important for high-throughput screening with density functional theory (DFT) for transition metal complexes, where incorrect charges can propagate substantial errors in computed electronic structure and energetics[35-37]. While calculations carried out by hand might be manually checked for erroneous charges, robust charge assignment is paramount in the proliferating codes designed for high-throughput or combinatorial screening workflows to construct and evaluate large libraries of theoretical complexes[38-41]. In the absence of reliable ligand charge assignments, DFT calculations, machine learning models trained on electronic descriptors, and automated workflows for generating novel complexes are prone to systematic and compounding errors[24,29,42,43].

Beyond structure and bonding, it is also useful to understand the context in which ligands have been applied experimentally. Many common ligands recur across literature, but not always for the same application area. Some ligands are tightly associated with a particular application area, such as biological[44-46] or photophysical studies[47-49], while others appear across multiple domains, including redox chemistry, and reactivity studies[50-54]. Understanding whether a molecule is specialized or broadly reused is valuable for rational design, where motifs or ligands associated with a field can be leveraged in screening protocols. Recent work has begun to address these challenges by utilizing existing workflows for classifying papers[55-59]. We recently curated the tmCAT, tmBIO, tmPHOTO, and tmSCO datasets of topic-classified TMCs[60] from the transition metal quantum mechanics (tmQM) data set[61]. While a promising foundation, extending these ideas to ligand-centric analysis requires workflows that accommodate multi-label behavior due to multiple uses of ligands and provide interpretable measures of application specificity. Such tools are particularly important for navigating large ligand libraries, where distinguishing between general-purpose ligands and those optimized for a specific application area can guide both experimental and computational screening efforts.

In this work, we address these challenges by constructing a large, experimentally grounded ligand dataset derived from 126,985 mononuclear TMCs in the CSD. From these structures, we identify 94,581 unique ligands, and we employ an iterative, weighted workflow to infer formal ligand charges across diverse coordination environments. This workflow enables consistent charge assignment for 66,810 ligands in our curated Boston Open-Shell Ligand (BOS-Lig) dataset. We then characterize ligand coordination chemistry by extracting coordinating atom identities, numbers, and experimentally observed binding-mode variability, allowing direct identification and classification of hemilabile behavior from crystallographic evidence. Finally, we associate ligands with functional application areas through bibliographic text mining, obtaining 38,115 ligands associated with publication titles/abstracts. From this set, we assign at least one application area label to 25,146 ligands, and we quantify the extent to which a ligand is specialized or broadly reused across the literature.

## 2. Results and Discussion

### 2a. Assigning Complex Charges

We first computed transition metal complex charges to then further assign individual ligand charges following a procedure first introduced in Ref. 30. We curated our initial set of TMCs from the March 2024 update of the Cambridge Structural Database (CSD) from the Cambridge Crystallographic Data Centre (CCDC).[25] Entries were restricted to those containing exactly one transition-metal atom (i.e., excluding lanthanides and actinides) in at least one structure, yielding 599,180 entries. Polymeric networks, disordered fragments, non-molecular species (i.e., entries with components that extended through each unit cell) were excluded (Supporting Information Text S1). Furthermore, entries with missing or undefined atomic sites were removed, ensuring that every retained structure corresponded to a molecule. Potential missing hydrogen atoms were added using the CCDC Python API, which infers absent hydrogens from the stored bond topology. For each structure, atom counts before and after hydrogen addition were compared, and complexes where the hydrogen count increased by more than the expected number were marked as "non-trustworthy" and excluded (Supporting Information Text S2). Overall, 83% of mined complexes required no additional hydrogens, 8% had suitably added hydrogens, and the remainder had hydrogen additions that were deemed unlikely to be correct.

We then decomposed each unit cell into its components, including TMCs, solvent molecules, and counterions, exporting each into a separate mol2 file. We computed a Weisfeiler–Lehman (WL) graph hash[62] for each component to identify each unique species. These hashes were collected automatically into a species dictionary, which tracked each unique molecular hash as well as the number of repeats for each hash. Because the WL hash depends only on molecular topology and atom identity, it does not distinguish between different charge or oxidation states of the same composition. This procedure yielded 467,853 unique molecular components in total. Although some TMCs include user-assigned net charges in their metadata, many do not, motivating us to develop a workflow to infer complex charges. A machine learning model such as cell2mol[34] could, in principle, be used to predict complex charges. We did not pursue this initially because cell2mol[34] fails to run on many structures in the CSD. These failures can typically be attributed to cell2mol having trouble parsing poorly resolved hydrogens, finding no possible charge assignment, and failing to reconstruct cells. We analyzed cell2mol failure cases and found some trends with respect to metal identity and complex size but could not resolve these errors (Supporting Information Figure S1). Thus, rather than solely relying on cell2mol, we instead implemented an iterative procedure. First, we started by manually defining the charges of 267 unique non-metal containing molecules that appear 50 or more times (i.e., common counterions, Figure 1 and Supporting Information Table S1). Then, we enforced unit cell neutrality, inferring charges of unknown unique components by subtracting the charge of known components from the unit cell. As this approach solves for more components, the dictionary containing assigned charges is iteratively increased.

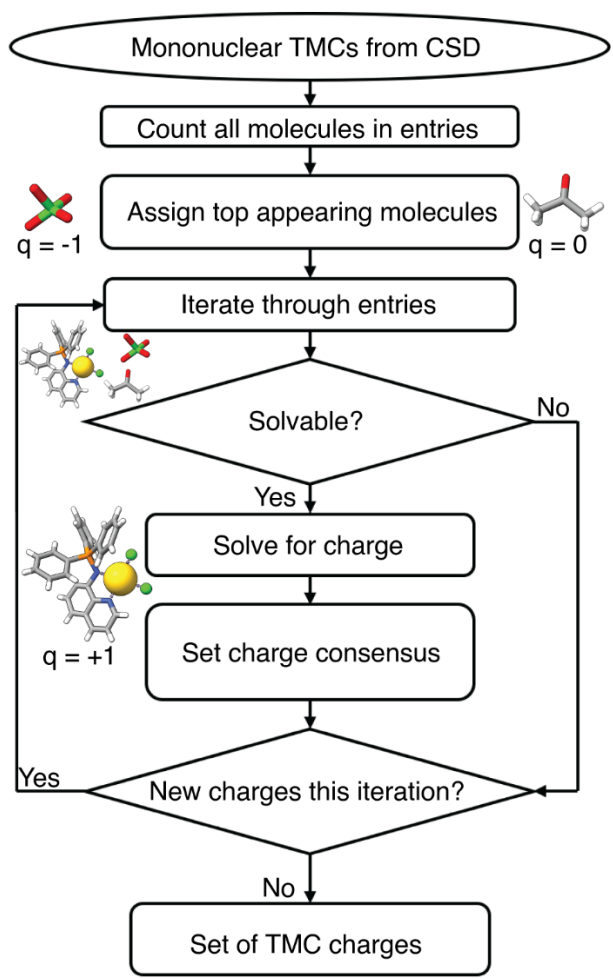

**Figure 1.** Iterative workflow for complex-level charge inference. Mononuclear TMCs extracted from the CSD are decomposed into molecular components and tallied to identify the most frequently occurring species. Initially defined seed charges enable iterative charge solving across entries under the assumption of unit-cell neturality. Each pass identifies solvable entries, assigns charges by difference, and updates the charge dictionary by majority consensus. The cycle repeats until convergence—defined as a full pass without new solvable species—after which non-TMC components are filtered out. An example is given with CSD refcode WIZJAS. Atom colors: carbon (gray), oxygen (red), nitrogen (blue), hydrogen (white), and chlorine (green), gold (yellow).

If different unit cells led to different charge assignments for entries identified to be identical based on their hash, the most commonly assigned charge was retained. This consensus procedure was applied iteratively across multiple inference passes. In the first pass, 4,622 molecules exhibited conflicting assignments across unit cells. Across the 4,622 conflicts in the first pass, the median number of total observations was 6, suggesting most disagreements came from cases with a handful of occurrences. To quantify the level of agreement across observations for a given formula, we defined a charge purity metric, equal to the percentage of observations supporting the most common assigned charge. We finalized the charge of a molecule only if it had at least three independent observations and a clear majority, i.e., estimated charge purity $\geq 67\%$. Molecules with fewer than three observations or lower charge purity were deferred for later passes (Supporting Information Text S3). Fractional charge predictions were rounded to the nearest whole integer while predictions ending in 0.5 were explicitly rejected to prevent spurious rounding. We repeated the workflow until no new charge assignments were obtained, requiring nine iterations in total (Supporting Information Table S2). By the end of the ninth iteration, 2,948 complexes were not assigned a charge due to a fractional value ending in 0.5, while 578 had non-integer fractional charges that were successfully rounded to an integer value (Supporting Information Text S4).

We next compared the resulting charges to CSD net charge annotations. Over this set, we observe agreement between inferred and CSD-reported charges was observed in 91% of cases (Supporting Information Figure S2). Examples of our iterative scheme being incorrect arose primarily from partial occupancies, duplicate ions, or missing counterions in the deposited structure (Supporting Information Figure S2). The CCDC's charge scheme makes assumptions about a complex's oxidation state, which can lead to incorrect charge assignments in our iterative scheme. An example of this is the complex with CSD refcode MBZPNI, a porphyrin-containing nickel complex. The CSD assumes an oxidation state of +3 on the nickel, resulting in a net charge of +1 for the TMC. However, when referring to the original publication, it is clear that the oxidation state should be +2 (Supporting Information Figure S2). To ensure an accurate initial set of complex charges, we only assigned charges where the CSD charge scheme and unit cell balancing approach agreed.

In order to next assign constituent ligand charges to each TMC, we also required assignment of the metal oxidation state. To do so, we used the text-based parsing approach from our prior work, only retaining cases with a single, unambiguous metal oxidation state, resulting in 113,934 oxidation states associated with unique TMCs.[30] We obtained additional oxidation states by using cell2mol[34] on a set of 43,625 complexes for which we had no prior known oxidation state. This resulted in 13,051 additional oxidation states (Supporting Information Text S5). After filtering only for complexes with a determined oxidation state (i.e., annotated or from cell2mol), we obtained a final dataset comprised of 126,985 unique mononuclear transition-metal complexes (Figure 2a). In our set, we have complete coverage of the 29 metals in all three rows of transition metals. The distribution of complex charges spans from –6 to +6, centered around neutral and moderately charged species (±2) (Figure 2b). Oxidation-state distributions across metals show expected trends, such as Fe, Co, Ni, and Cu being predominantly +2 while Ru and Ir are mainly +3, suggesting that the charge and oxidation state assignments are likely reasonable (Figure 2c).

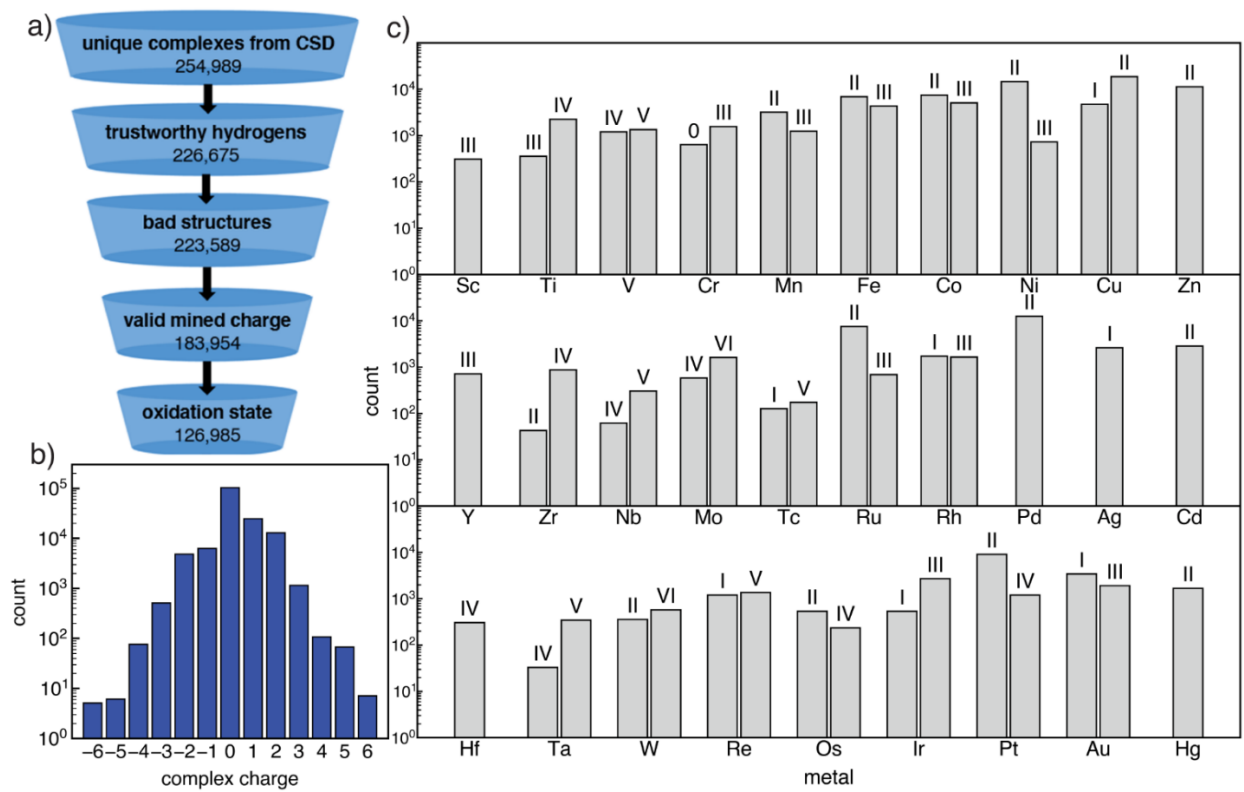

**Figure 2.** Count, charge and oxidation state distributions for TMCs. a) Funnel diagram illustrating the filtering of TMCs from the CSD. Starting from 254,989 unique complexes, structures were screened for hydrogen completeness, structural validity, and the presence of a consistent complex-level charge and oxidation state, yielding 126,985 high-confidence entries suitable for ligand charge analysis. b) Histogram of assigned complex charges across the dataset. Most complexes are neutral or carry a small integer charges ($-2$ to $+2$). c) Oxidation state distributions for each transition metal, showing the relative abundance of the most common states. The expected dominant oxidation states (II for Ni, Cu, Pd; III for Co, Fe; VI for Mo and W) are well represented.

## 2b. Assigning Ligand Charges

With complex-level charges and metal oxidation states established, we next sought to assign charges for constituent ligands. For each transition metal complex with a known total charge and metal oxidation state, the residual charge must be assigned to its ligands. Because our dataset contains many assigned complex charges, these constraints enable robust ligand-charge estimates from consensus, as an expansion of the approach introduced in our prior work.[30] For each TMC, we first isolated individual ligands by decomposing the molecular graph into a set of connected, non-metal fragments along with the central metal atom and assigning each resulting ligand a molecular graph hash. The hashing scheme distinguishes ligands only by their underlying chemical connectivity and atom identities (Supporting Information Text S6). For each unique ligand hash, we recorded the number of occurrences and the crystallographic R-factor of the total unit cell, which we used in later steps for weighting observations and consistency checks. The distribution of ligand frequencies shows both the large number of unique ligand graphs and the sharp disparity between rare and frequently used ligands (Figure 3).

Although most ligands appear only once or a few times, a small subset occurs extremely frequently across the dataset (Figure 3). Following the approach introduced by Duan et al.[30], we initiate ligand-charge assignment with homoleptic complexes containing these high-frequency ligands because the ligand charges in these TMCs can be determined by the total charge divided by the number of ligands (Figure 4). We use an iterative extension of this procedure for heteroleptic complexes. Whenever a complex contains only ligands whose charges have already been assigned (e.g., from homoleptic TMCs), we infer charges for all remaining ligands in the TMC. Each new assignment updates the global dictionary of ligand charges, enabling additional heteroleptic cases to be resolved in subsequent passes. We repeat this iterative procedure until complete review of the set of TMCs yields no further assignments (Figure 4a).

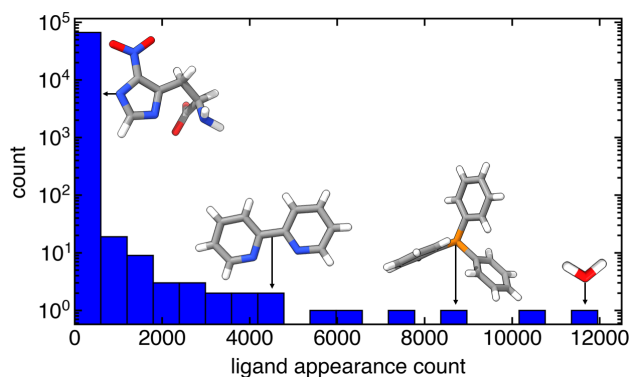

**Figure 3.** Distribution of ligand appearance frequencies across the curated mononuclear TMC dataset. Most ligands occur only once or a few times (52,944 ligands appear 3 or less times, corresponding to the tallest bin at the left in this histogram), resulting in a long-tailed frequency profile. A small subset appears hundreds to thousands of times, while the majority occur infrequently. Atom colors: carbon (gray), oxygen (red), nitrogen (blue), phosphorus (orange), and hydrogen (white).

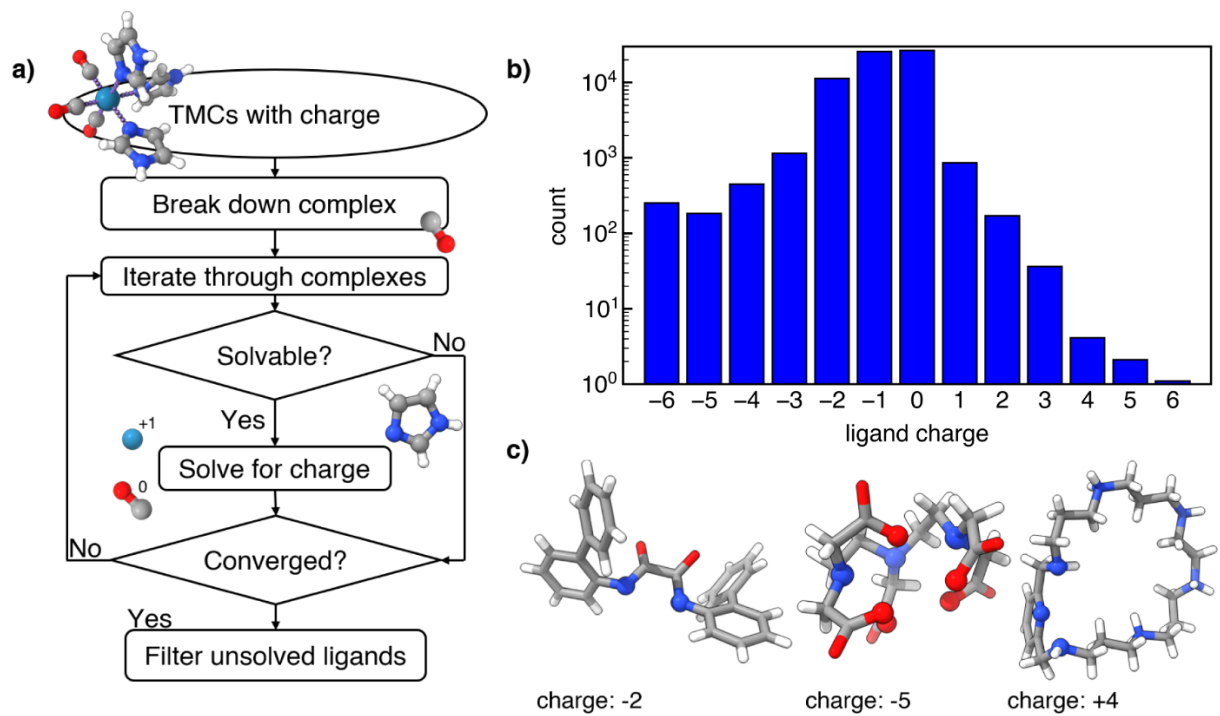

**Figure 4.** Ligand charge workflow and results. a) Flowchart of the iterative procedure used to assign ligand charges. b) Distribution of the assigned ligand charges. c) Representative ligand examples illustrating typical and out-of-range assignments: a ligand with a charge of −2 (within the main distribution), alongside cases of rarely assigned charges (−5 and +4) outside of the main distribution. Atom colors: carbon (gray), oxygen (red), nitrogen (blue), hydrogen (white), and rhenium (cyan).

One challenge is that different TMC depositions might lead to deduction of different charges for two instances of the same ligand (i.e., as judged by graph hash connectivity). To resolve these conflicts, we introduce a weighted procedure that ranks each candidate charge based on a judgement of the quality of its CSD refcode source. That is, we assign higher weight to complexes from: i) well-resolved structures (i.e., lower R-factors), ii) with fewer unique ligand types, and iii) where the charge assignment deduced is an integer (Supporting Information Text S7). We first record all possible charge assignments, and, at the end of each iteration, we assign the charge with the highest accumulated weight from all observations (Figure 5). During the iterations over all TMCs, a ligand may receive a provisional charge early on that may be updated once additional data from other ligands is incorporated.

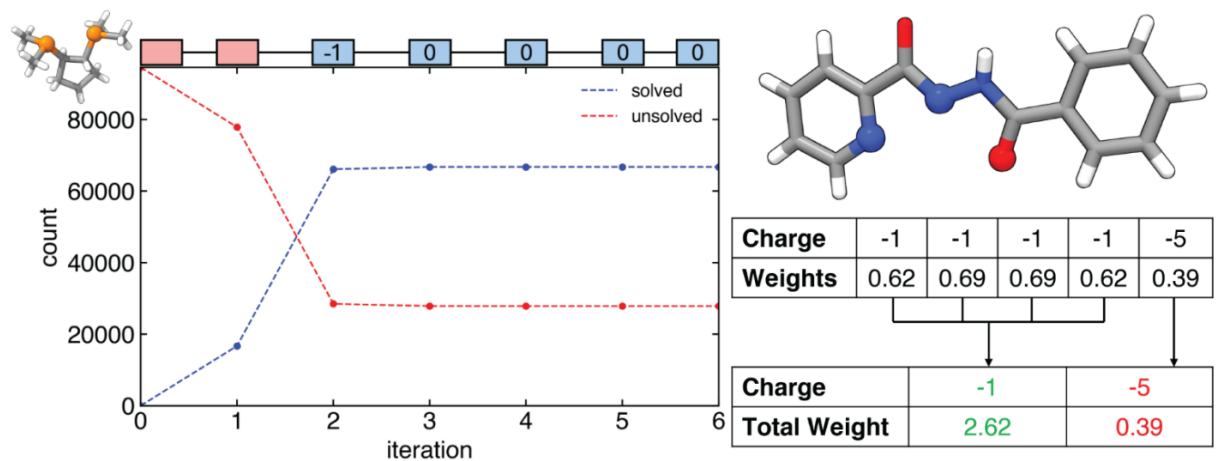

**Figure 5.** Weighted ligand charge assignment. (Top) An example inconsistent ligand whose assigned charge transitions from –1 to 0 after the second iteration. Boxes denote the ligand's charge in each iteration, with color indicating whether the ligand was solved (blue) or unsolved (red) at that stage. (Left) Scatter plot depicting the evolution of the number of solved and unsolved across iterations. (Right) An example ligand highlighting the weighted-vote scheme used to determine the charge assignment. Individual charge estimates and their weights are listed, along with the resulting weighted totals used to select the final assigned charge. Atom colors: carbon (gray), oxygen (red), nitrogen (blue), and hydrogen (white).

Using the weighting procedure for charge assignment, the workflow for ligand charge assignment converged in five iterations (Supporting Information Table S3). Across the curated complexes, we identified 94,581 unique ligands and successfully mined charges for 66,810 of them, corresponding to 71% of the ligand chemical space (Supporting Information Text S8). Ligand charges predominantly ranged from –4 to +2, with neutral ligands being the most common, accounting for 39.7% of assigned ligands, followed by ligands with charges of –1 (38.5%) and –2 (17%) (Figure 4b). Some values fall outside of the typical -4 to +2 net charge range, albeit rarely (0.01% of all assignments), and in many cases still appear realistic (i.e. within ± 6 net charge value, see Figure 4).

Across the dataset, ligand net charges exhibit strong, systematic clustering by functional group motifs. Neutral ligands are dominated by hydrocarbon frameworks and common nitrogen-containing ligands such as pyridines, imidazoles, and chelating diimines, exemplified by 2,2′-bipyridine. These ligands span a wide range of sizes and number of metal-coordinating atoms yet are overwhelmingly neutral. In contrast, negatively charged ligands are strongly enriched in oxygen-based motifs, which introduce formal anionic character even for relatively small ligands. Carboxylates and phenoxides are most frequently assigned –1, while ligands containing multiple anionic donor groups, including dicarboxylates, sulfonates, and phosphate-derived ligands, are predominantly assigned –2, accounting for approximately 30% of all negatively charged ligands. Porphyrins (e.g., tetraphenylporphyrin) constitute a prominent and chemically distinct class within the dataset and are consistently labeled as –2. Positively charged ligands are comparatively rare and are largely confined to nitrogen-rich cationic or protonated species, such as protonated tris(pyridylmethyl)amine derivatives. Ligands assigned charges outside the common –4 to +2 range are rare and primarily reflect highly polyanionic species, such as triphosphate, or multiply protonated ligands.

To quantify the uniformity in our assigned charges, we defined a charge inconsistency score, which is the ratio of unique charges to total calculated charges, to measure the spread among the raw charge estimates for each ligand (Supporting Information Figure S3). Here, a ligand is considered *confidently consistent* when all observed charge assignments agree, corresponding to an inconsistency score of zero. Ligands with multiple observations but only one unique assigned charge are therefore treated as fully consistent, while any ligand with more than one unique assigned charge is classified as potentially inconsistent. Ligands with low inconsistency scores, notably chlorine (inconsistency = 0.0004), the most commonly appearing ligand in the CSD, correspond to high-confidence assignments, as few examples deviate from the overwhelming consensus of –1. Ligands with broad or multimodal distributions were treated cautiously, such as 5,10,15-tris(2,4,6-(triphenyl)phenyl)corrole, the most inconsistent ligand with more than 10 appearances (inconsistency = 0.42). It is worth noting that despite the inconsistency, the correct charge of –3 was the final extracted charge due to its plurality, as the deviating charges were split among multiple values without clear consensus. In total, 2,640 ligands (~4%) were classified as potentially inconsistent (i.e., having more than one charge assignment). After identifying ligands with inconsistency scores above 0.25 (i.e., ligands with multiple discrepancies), a 1,476 ligands subset (~2%) were identified that exhibited more substantial disagreement among their assigned charge values. This low occurrence of our final assignment of the charge as inconsistent (1,476) is particularly noteworthy considering 21,187 ligands had multiple initial determinations of charge. This indicates that final charge assignments were largely consistent in the vast majority (i.e., 19,711 or 21,187 or 88%) of these ligands. The charge assignments are confidently consistent for 65,383 ligands in the dataset, highlighting the lack of conflicting charge assignments and aiding in overall confidence for charge calculations.

We next compared our inferred ligand charges to prior work. Duan et al.[30] first introduced an iterative workflow "OctLig" that assigned charges to 8,805 ligands extracted from 28,006 octahedral CSD complexes. In contrast, our approach applies the same overall logic to a far larger and chemically broader set of mononuclear TMCs (i.e., beyond octahedral TMCs), yielding charge assignments for nearly an order of magnitude more ligands as a result. A recently introduced workflow, DART[32], assigns ligand charges by treating the entire dataset of TMCs as one system of equations, where they then solve a single global least-squares problem to find the set of ligand charges that best satisfies all complexes simultaneously. In total they report 41,018 confidently assigned ligand charges. Finally, the tmQMg-L dataset[29] provides ca. 30,000 ligands obtained from CSD structures with charges derived from NBO-based Lewis structures. While it spans a wider range of coordination motifs than OctLig, its overall ligand coverage is still roughly half that achieved in this work. In a related approach, the ReaLigands set was generated by predicting charges for more than 30,000 ligands from mononuclear CSD complexes and then assigning ligand charges using a random-forest classifier trained on GFN2-xTB-derived features (i.e. HOMO–LUMO gap, internal force, and SCF convergence behavior) evaluated across candidate charge states.[33] As in the other cases, ReaLigands spans a much more limited chemical space compared to the BOS-Lig dataset provided here. Thus, our set of ligands represents the largest set extracted to date from experimentally characterized structures.

To assess the accuracy of our approach, we benchmarked it against two alternative approaches: i) cell2mol's ligand-charge predictions[34] and ii) an octet-rule heuristic (Figure 6). For a randomized subset of ligands, we manually inspected assignments to provide ground-truth validation. We took a random subset of 7,309 complexes that correspond to 4,000 unique ligands. Among that set, cell2mol successfully assigned charges on 4,280 TMCs, allowing for the extraction of 1,743 ligand charges. Most ligands (i.e., 1,674) had charges that matched our charge calculations, and differences slightly favored cell2mol's predictions being correct (i.e., 33 of 69 charges being correct as opposed to our method being correct in 30 of 69) upon manual inspection of the cases that disagreed (Figure 6). Nevertheless, while cell2mol is accurate, it often fails to complete a charge assignment on diverse or structurally complex TMCs, excluding large portions of the chemical space that our approach can capture.

We then compared our charge assignment schemes to a more universally applicable, simpler method, valence-based electron counting. Across all assigned ligands, our charge assignments agreed with octet-rule predictions 87% of the time. For 40 randomly selected ligands with disagreements, we found that our workflow produced the correct charge in 28 cases, while the octet rule provided the correct charge in eight, and both were ambiguous in four cases (Figure 6). The octet-rule heuristic is broadly applicable, but it performs poorly for $\pi$-acceptor ligands, delocalized systems, and hypervalent donors. Thus, the iterative workflow provides benefits over both octet rule and cell2mol assignments. These results highlight that the iterative scheme provides chemically sensible ligand charges, particularly in bonding environments where simpler heuristic methods fail.

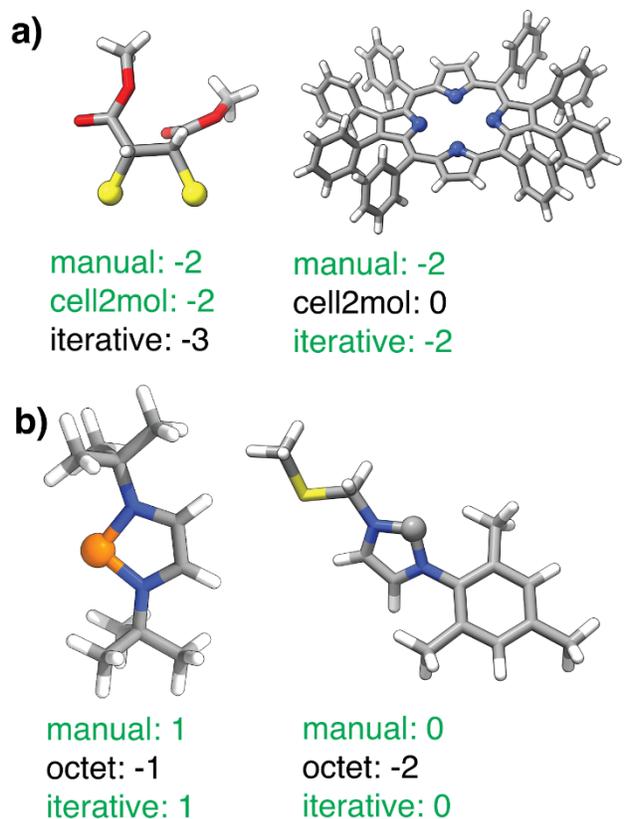

**a)**

manual: -2
cell2mol: -2
iterative: -3

manual: -2
cell2mol: 0
iterative: -2

**b)**

manual: 1
octet: -1
iterative: 1

manual: 0
octet: -2
iterative: 0

**Figure 6.** Comparison of iterative ligand-charge assignments with alternative charge-assignment schemes. a) Representative ligands where the iterative workflow disagrees with cell2mol. b) Representative ligands where the iterative workflow disagrees with the octet rule, showcasing an hypervalent phosphorous containing ligand (double bonds to the central metal), and an n-heterocyclic carbene containing ligand. Atom colors: carbon (gray), oxygen (red), nitrogen (blue), hydrogen (white), phosphorous (orange), and sulfur (yellow).

## 2c. Ligand Coordination Chemistry

To characterize how ligands coordinate metal centers across the dataset, we extracted the donor atoms associated with each observed binding mode. Across the dataset, the average number of atoms of our ligands coordinating to the metal corresponds to 2.77 donor atoms per binding mode, with the distribution dominated by two (24,908) and one (15,158) coordinating atom cases but with substantial populations of ligands with three (12,245) and four (9,746) coordinating atoms (Figure 7a). Most ligands exhibit nitrogen-based coordination (42.7% of all donor sites), followed by carbon (29.7%) and oxygen (14.6%), with phosphorus (5.7%) and sulfur (5.4%) together comprising an additional ~11% of donor atoms. A chemically diverse long tail of less common donors (e.g., Se, Si, As, B, halides) is also present but makes up a small percentage of the total ligand space (Figure 7b).

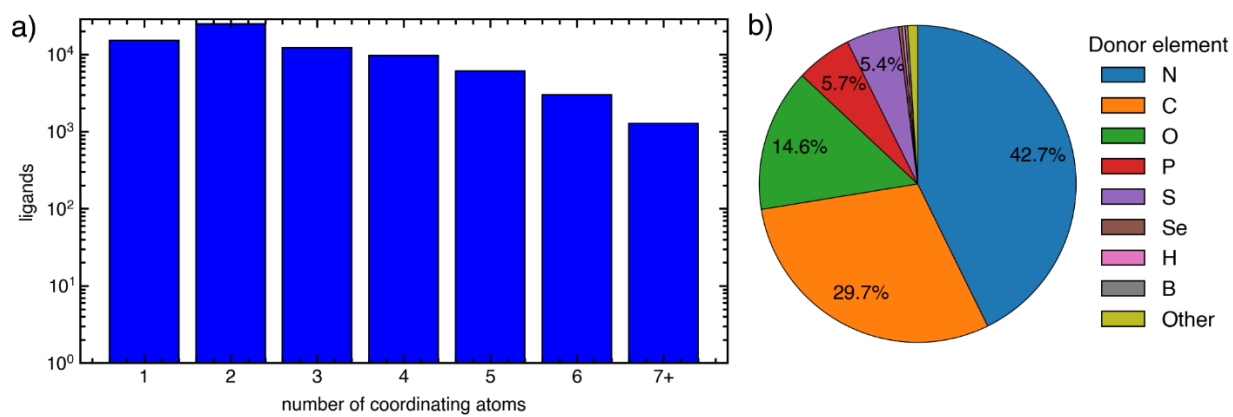

**Figure 7.** Ligand coordination details. a) Distribution of number of ligand coordinating atoms across all binding modes in the dataset. b) Elemental composition of donor atoms, expressed as the percentage contribution of each donor element.

We further classify the charge-assigned ligands as hemilabile using our recently introduced four-type hemilability categorization.[63] Type 1 and 2 hemilability both involve changes in coordination number, with type 1 corresponding to donor loss or gain where the lower-coordination mode coordinating atoms are a subset of the higher-coordination mode coordinating atoms, and type 2 involving coordination-number changes between non-subset donor sets of atoms. Type 3 and 4 hemilability preserve coordination numbers and differ by donor atom overlap, with type 3 exhibiting partially overlapping coordinating atoms and type 4 involving mutually exclusive donor sets. Comparing the 6,480 pairs of hemilabile ligands of any type in our set to the curated hemilabile set we reported in Ref. 63, we similarly observe a high percentage of type 1 ligands (here, 67.2%) and a similarly small minority of type 3 (5.8%) and type 4 (4.8%) ligands but a higher contribution from type 2 hemilability (22.2%). This shift (6.9% in the original work vs. 22.2% here) is likely due to subtle differences in allowed number of coordinating atoms (here, no maximum is set), and we do not exclude crystal structures that may have errors, leading to higher numbers of predicted hemilabile modes.[63]

## 2d. Assigning Ligand Application Area

Beyond ligand structure, identifying the application area of a ligand is a useful next step leveraging large experimental datasets for TMC design. Prior applications in the literature provide a powerful basis for guiding screening efforts. Recently, we demonstrated[60] this concept by associating TMCs in the tmQM[61] dataset with application-level topics inferred from publication text, and we now extend this approach to annotating our larger set of ligands.

We employed a revised protocol from the original work i) to allow for ligands to belong to multiple categories and ii) to more flexibly define categories. As in prior work, we carry out topic assignment from the title and abstract of the paper associated with each CSD refcode. Bibliographic records were obtained for 157,874 complexes and 43,744 unique ligands, corresponding to 59.7% of crystallographically characterized TMCs and 65.5% of ligands (Figure 8a and Supporting Information Text S9). We adopted a BERTopic-based[55] workflow as in prior work[60], with select modifications (Supporting Information Text S10). Using this approach, we curate application-linked ligand subsets for the following application areas: reactivity and catalysis (react), biological (bio), spin-state and magnetism (magnet), redox chemistry (redox), and photophysical (photo) chemistry. Of the 43,744 ligands with bibliographic text coverage, 25,146 could be assigned at least one application area label. A slight majority (i.e., 12,751 or 50.7%) appear in one application area, with other ligands appearing in two or more areas (i.e., 12,395 or 49.3%) and a smaller but substantive subset of these (i.e., 3,348 or 13.3%) appear in three or more. A small number of ligands 1,335 ligands (5.3%) appear in four areas and 422 (1.9%) appear in all five areas (Figure 8b). The 'react' and 'magnet' application areas had the most overlap with the other application areas, whereas 'bio' and 'photo' were more self-contained (Figure 9). React emerged as the most populated application area, dominated by neutral and monodentate ligands, modulating metal reactivity through single-site donor interactions. Meanwhile photo and bio were the smallest, most non-overlapping areas, whereas redox and magnet overlapping with many of the other areas. Even though ligands may appear in multiple application areas, they may be more frequently associated with just one of the application areas.

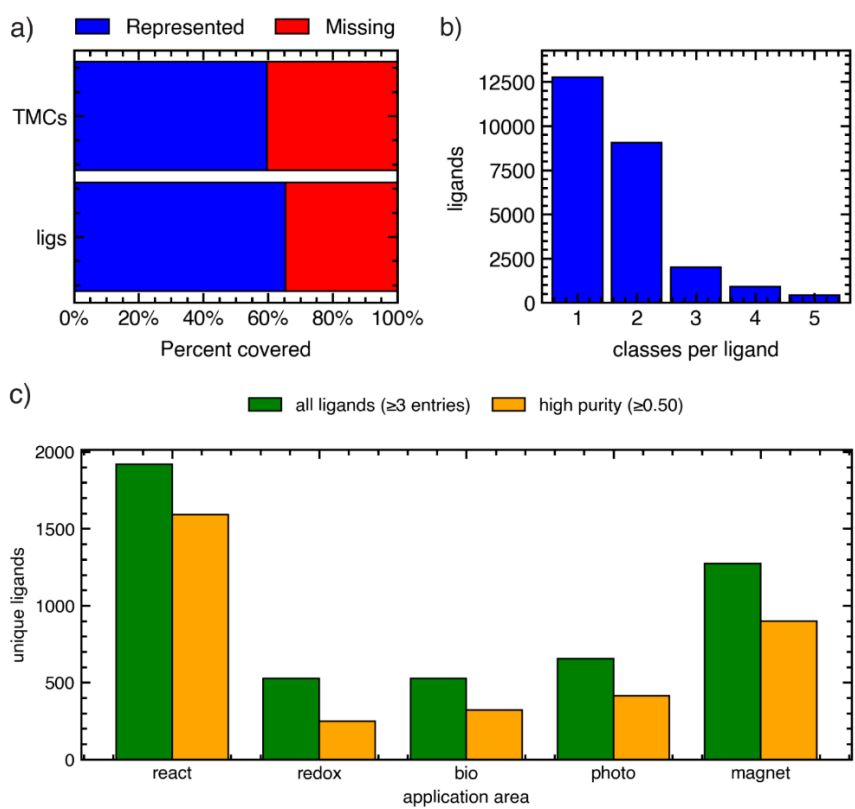

**Figure 8.** Coverage and application breadth of ligands and transition-metal complexes with associated bibliographic text. a) Fraction of crystallographically characterized transition-metal complexes and unique ligands that could be classified, relative to those lacking text coverage. b) Distribution of application breadth, defined as the number of distinct application classes in which a given ligand appears. c) Unique ligands per application area (fraction of ligand in each application area with respect to that ligand in all application areas) with 3 or more CSD entries compared to the subset exhibiting high (≥0.50) application purity.

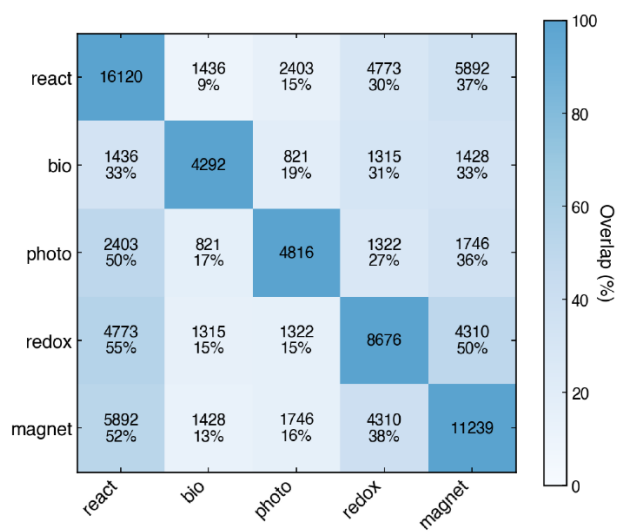

**Figure 9.** Pairwise overlap of application areas for ligands. Each cell reports the number of ligands shared between two application areas and the percentage of ligands in the row category that also appear in the column category. Due to the difference in size of the two classes, the percentage will differ in the upper and lower triangle of the matrix. Diagonal entries correspond to all ligands associated with that category.

To determine whether a ligand has a dominant application area, we define an application area purity ($p$):

$$p_{l,a} = \frac{n_{l,a}}{\sum_{a' \in A} n_{l,a'}}$$

$$A = \{\text{react, bio, magnet, redox, photo}\}$$

where $n_{l,a}$ is the number of times a ligand $l$ falls under application $a$ (Supporting Information Text S11). Despite overlap of 'react' ligands with other classes, the class has many notable high-purity ($p > 0.5$) ligands, such as triphenylphosphine (react $p = 0.60$), which are frequently used in homogeneous transition-metal catalysis[64,65], hydride and oxidative-addition chemistry[66,67], and mechanistic studies of ligand dissociation and lability[68]. In contrast, high-purity 'bio' ligands are generally amine-rich, including phenylalaninato (bio $p = 0.82$, i.e., anionic phenylalanine), which is associated with DNA[69,70] and RNA[71] binding and protein interactions[72]. Many of the high-purity, bio-classified ligands have been employed in cytotoxic TMCs, particularly those featuring copper centers[69-73]. Ligands with high purity from the photo application area predominantly correspond to two types. Many, such as difluoro-2-(2-pyridyl)phenyl (photo $p$ = 0.83), have CF₃-substituted aryl motifs, useful for blue-shifted phosphorescence and high electroluminescent efficiency[74-76]. Others contain bulky diphosphines, such as (oxydi-2,1-phenylene)bis(diphenylphosphine) (photo $p$ = 0.67), which serve to rigidify the metal environment[77,78] and promote bright, long-lived emission in light-emitting devices[77-80]. Even though redox and magnet application areas exhibit high overlap, there are representative ligand cases of high purity. For redox, this includes notably 1,2-di(2,6-diisopropylphenylimino)acenaphthene (redox $p$ = 0.55), a redox-active α-diimine ligand[81,82] for ligand-centered electron transfer[83]. For magnet, a representative example is isothiocyanide (sco $p$ = 0.76), which is used as an ligand for spin-crossover[84-86], magnetic anisotropy[87-90], and single-ion magnet systems[91-93]. The fact that high purity ligands correspond to ones we expect to reside in the single sets is an encouraging signal that lower purity occurs ligands are indeed used in more than one application area rather than a reflection of the failure of the application area classification. For example, the polypyridyl ligand 4,4′-di-tert-butyl-2,2′-bipyridine (react $p$ = 0.42, photo $p$ = 0.37) appears both in organometallic reactivity studies[94,95] and in photochemical work on emissive Pt(II) complexes[96,97], and its low purity metrics reflects this. To study ligands specifically for a single application area in a high-throughput computational or experimental screen, a user can choose to prioritize ligands that are confidently classified (i.e., high-purity) only in that application area.

## 2e. Sharing the BOS-Lig Dataset.

To share both ligand charge assignments and application areas with the broader chemical community, we have developed the BOS-Lig Browser, a browser-based platform for exploring ligand charge, observed coordination modes, and application-specific usage patterns directly from graph hash, chemical formula, or SMILES input (https://molsimplify.mit.edu/bos-lig, Figure 10). The BOS-Lig Browser enables users to visualize ligand structures and observed coordination modes, examine application assignments and purities across major use categories, and identify related ligands with similar application profiles. By making these data and analyses directly accessible through an interactive web interface, the BOS-Lig Browser provides an intuitive resource for navigating ligand behavior in transition metal chemistry without requiring computational expertise.

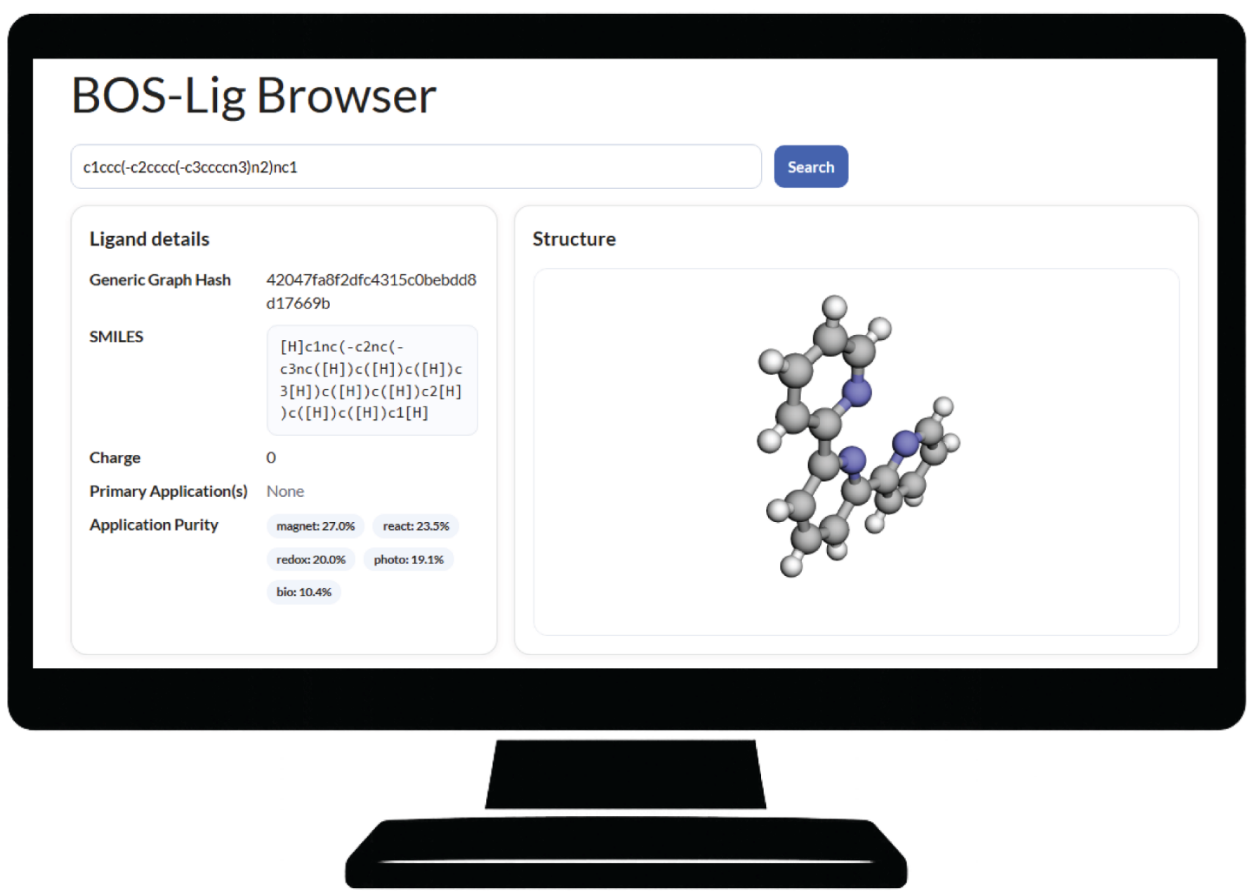

**Figure 10.** Screenshot of the BOS-Lig Browser web interface, a browser-based platform for exploring ligand charge, observed coordination modes, and application-specific usage patterns from graph hash or SMILES input. The interface allows users to query a ligand, visualize its structure, inspect key ligand properties including charge and application purity, and identify related ligands with similar application profiles.

## 3. Conclusions

In this work, we developed a workflow for extracting ligand information from crystallographic structures of transition metal complexes to enable high-throughput screening and data-driven ligand design. Starting from 126,985 mononuclear complexes curated from the Cambridge Structural Database (CSD), we constructed a ligand-centered dataset containing 94,581 unique ligands. We curated the BOS-Lig dataset by assigning net charges for 66,810 of these ligands using an iterative consensus-based procedure. This workflow combined unit-cell charge balancing, oxidation-state information, and weighted voting across crystallographic observations to determine ligand charges across diverse coordination environments. Compared to prior approaches, this strategy expanded the number of experimentally derived ligand charge assignments by nearly an order of magnitude while maintaining agreement with alternative charge assignment methods (e.g., octet rule) where consistency was expected. We also introduced a purity metric to evaluate when charge assignments might be sensitive to conflicting reports of the ligand in the CSD. Analysis of the donor atoms and number of coordinating atoms in these ligands revealed approximately 8% of ligands exhibit multiple coordination modes. This combined information about ligand charge and preferred coordination is expected to be useful for guiding *de novo* TMC generation from constituent ligands.

In addition to ligand analysis based on structure, we extended a previously introduced topic-modeling workflow to associate ligands with application areas. This assignment was successful for 25,146 ligands in our set. We also allowed for ligands to be associated with multiple application areas and introduced an application purity metric to quantify whether ligands were specialized or reused broadly across different areas of transition metal chemistry. Together, our datasets provide key information about charge, coordination, and application area. The resulting dataset provides experimentally grounded information that is expected to be useful in constructing new complexes for computational screening, improving machine-learning models for transition metal chemistry, and guiding data-driven ligand selection. All datasets and analysis workflows are made publicly available via the BOS-Lig webserver, https://molsimplify.mit.edu/bos-lig, and Zenodo [98], enabling further data-driven exploration of ligand chemistry.

ASSOCIATED CONTENT

**Supporting Information.**

Method for TMC dataset construction from the CSD, including procedures for identifying transition metal complexes and filtering polymeric and disordered entries; component-level mining of CSD entries and metadata extraction; procedures for hydrogen addition, ghost-atom detection, and structural trustworthiness assessment; statistics on the most common molecular species present; description of the iterative unit-cell charge balancing workflow used for complex charge assignment and propagation of charge information across crystallographic entries; statistics on solved component charges across iterations and analysis of fractional charge cases and rounding procedures; comparison between iterative charge assignments and deposited CSD charges for representative complexes; workflow for mining metal oxidation states using cell2mol and integration of these results with oxidation states extracted from CSD metadata; graph-based procedures for isolating ligands from complexes through removal of metal centers and partitioning of molecular graphs into ligand fragments; calculation of graph hashes for ligand identification across complexes; statistics on solved ligand charges across iterations and coverage of the solvable ligand chemical space; distributions and analysis of ligand charge inconsistency scores; procedures for constructing a bibliographic corpus associated with CSD entries using DOI metadata with the Web of Science API; details of topic-model seeding and classification for ligand application areas including reactivity, redox chemistry, biological, photophysical, and magnetism applications; and definition of the ligand topic-purity metric used to quantify the selectivity of ligands toward specific functional domains. (PDF)

**Data and Software Availability statement**

All data required to reproduce this work is provided either in the supporting information PDF file or in the attached Zenodo. Interactive access to the dataset can be found via https://molsimplify.mit.edu/bos-lig.

AUTHOR INFORMATION

**Corresponding Author**

*email:hjkulik@mit.edu

**Author Contributions**

Roland G St. Michel: Conceptualization, Data curation, Formal Analysis, Investigation, Methodology, Software, Validation, Writing – original draft, Writing – review & editing; Ryan J Jang: Data curation, Formal Analysis, Investigation; Aaron G Garrison: Data curation, Investigation, Methodology, Software, Writing – review & editing; Ilia Kevlishvili: Formal Analysis; Heather J Kulik: Conceptualization, Funding acquisition, Project administration, Supervision, Writing – review & editing

**Notes**

The authors declare no competing financial interest.

ACKNOWLEDGMENTS

Funding was provided by a UPI from The Dow Chemical Company. A.G.G was partially supported by Chemical Engineering MathWorks Engineering Fellowships. H.J.K. is supported by a Simon Family Faculty Research Innovation Fund and an Alfred P. Sloan Fellowship in Chemistry. The authors acknowledge the MIT SuperCloud and Lincoln Laboratory Supercomputing Center for providing HPC resources that have contributed to the research results reported in this work.

**For Table of Contents Use Only**

| Mine Entries | Assign Lig. Charge | Classify |
|---|---|---|
| +1    126,985 | 66,810 | 25,146 |

+1

-1

+0

-2

bio

photo

Roland G. St. Michel[1], Ryan Jang[2], Aaron G. Garrison[3], Ilia Kevlishvili[3], and Heather J. Kulik[2,3,*]

[1]*Department of Materials Science and Engineering, Massachusetts Institute of Technology, Cambridge, MA 02139, USA*

[2]*Department of Chemistry, Massachusetts Institute of Technology, Cambridge, MA 02139, USA*

[3]*Department of Chemical Engineering, Massachusetts Institute of Technology, Cambridge, MA 02139, USA*

*corresponding author email: hjkulik@mit.edu

**Contents**

**Text S1.** Filtering of Polymeric, Disordered, and Non-molecular Entries.

An initial set of CSD refcodes were obtained using ConQuest to find all refcodes containing a transition metal[1]. Then all CSD entries were examined component-by-component to identify transition metal complexes. Filtering was implemented primarily using molSimplify[2,3] and the CCDC Python API[1].

1. **Transition-metal and stoichiometric filters**

   Each crystallographic component was checked using `csd_entry.molecule.components`. To find transition metals, the element compositions of components were compared against a curated list of 30 transition-metal symbols. Only components containing exactly one transition-metal atom (`sum(metal_counts) == 1`) were accepted, eliminating both multinuclear clusters and organic molecules.

2. **Removal of polymeric and non-bonded components**

   If a component was the only transition metal containing component, the entry was rejected if `comp.is_polymeric == True`, eliminating coordination polymers and infinite lattice fragments. Components lacking any bonds (`len(comp.bonds) == 0`) were also discarded, ensuring that only connected molecular graphs were retained.

3. **Connectivity-based exclusion of fragmented or mis-bonded species**

   Each TMC was reconstructed as a molSimplify `mol3D` object to analyze the internal bond network. The metal coordination number (`len(orig_mol3D.getBondedAtomsSmart(orig_mol3D.findMetal()[0]))`) and the overall distribution of atomic connectivities (`orig_mol3D.graph.sum(axis=1)`) were inspected. Complexes were kept only when a single coordination center existed with between 2 and 12 neighbors and no additional high-coordination sites (i.e. more than sites with more 5 or more bonds).

4. **CSD entry component mining**

   For all TMCs that passed steps 1, 2, and 3, all components within the corresponding entry were saved and mined. Various CSD metadata including charge, chemical formula name, refcode, and refcode_plus, were recorded for each component, alongside a mol2 string representing the structure of the isolated component.

**Text S2.** Hydrogen Addition, Ghost-Atom Detection, and Structural Trustworthiness Assessment.

To ensure chemically complete molecular representations prior to charge and ligand analysis, each accepted CSD component underwent a structured hydrogen-addition and verification procedure implemented through the CCDC Python API[1].

1. **Baseline atom counts and ghost-atom screening**
   For each component, the total number of atomic sites and the number of properly labeled sites were recorded prior to any modification. Ghost atoms were defined as entries whose atom labels contained no numeric index (e.g., "X", or blank labels). We then determine the expected number of new hydrogens based on what would be needed to satisfy the valence of each atom in the existing bond network.

2. **Hydrogen addition with redundancy handling**
   Each component was duplicated, and hydrogen atoms were added using CCDC's valence-based hydrogen addition algorithm
   (`ccdc.io.EntryReader.add_hydrogens('missing')`). If hydrogen addition failed, primarily due to undefined bond order, the code attempted to infer bond orders and then attempted hydrogen addition for a second time.

3. **Post-addition consistency tests**
   After hydrogen addition, a binary flag (`h_trustworthy`) was stored for every complex, based on the results of the hydrogen addition:

   **Case A – No net change:**
   The hydrogen addition introduced no new atoms, confirming that all hydrogens had been properly represented in the CSD entry. These complexes were marked as trustworthy without modification.

   **Case B – Expected increase only:**
   The component added the expected number of missing hydrogens and introduced no ghost atoms or spurious sites. The hydrogen-added version replaced the original component (`self.csd_component = h_added_comp`) and was marked as trustworthy.

   **Case C – Over-addition or mismatch:**
   When the added atom count exceeded expectations, indicating insertion of extra, unlabeled, or ambiguous atoms, the structure was preserved only for recordkeeping. The suspect hydrogen-added representation was stored in the field `csd_mol2string_suspect_h_added`, while the main workflow reverted to the original component. These entries were then classified as non-trustworthy.

   **Case D – Complete failure:**

   If both addition attempts failed the component was flagged as non-trustworthy.

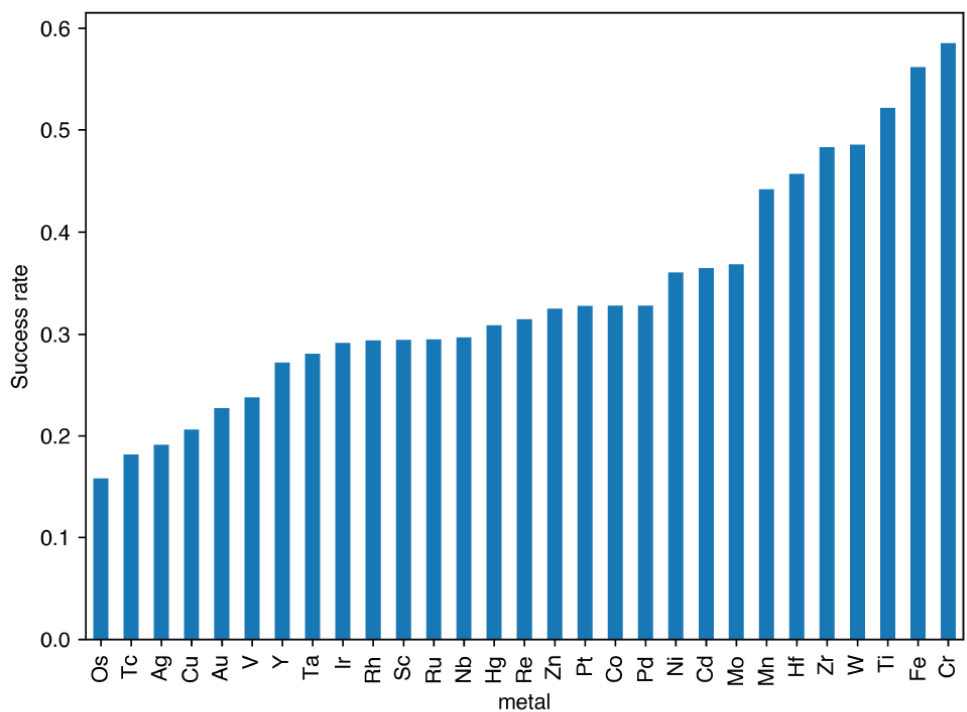

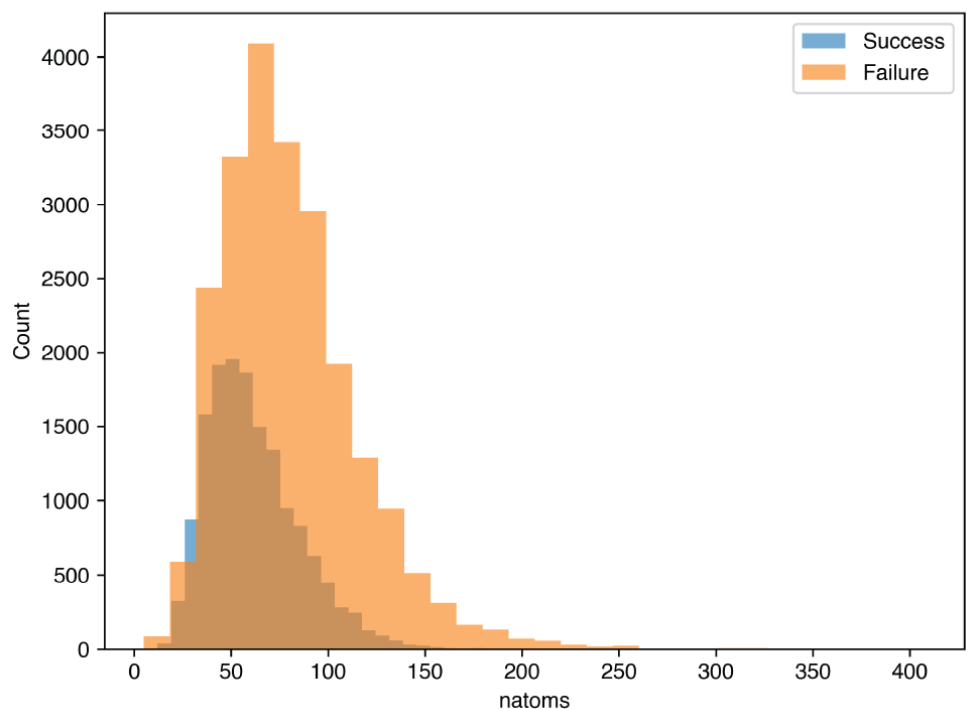

**Figure S1.** Failure analysis for cell2mol: (top) success rate of cell2mol with respect to different complex metal centers. (bottom) Distribution of the number of atoms for complexes that succeeded (blue) or failed (orange) to run with cell2mol.

**Table S1.** Top 20 occurring components within CSD entries containing a mononuclear TMC. Formula column has the atom identity, followed by count, and the CSD stated charge for the molecule at the end.

| formula | graph_hash | count | recursive_charge |
|---|---|---|---|
| H2 O1 | 2c96eb4288d63b2a9437ddd4172e67f3 | 81857 | 0 |
| C1 H2 Cl2 | 3bdeb5308de355e8960ff57bcef4c845 | 29918 | 0 |
| Cl1 O4 1- | f2193b01f90d729137d5242e55163a72 | 25123 | -1 |
| F6 P1 1- | a6c7e4cf1839b9796e83148fbc302e3a | 23193 | -1 |
| C2 H3 N1 | 10ca7fcdb78072eef40d96436ba108f9 | 18758 | 0 |
| B1 F4 1- | d9981e98e27e6bc05916add06369cdaa | 16130 | -1 |
| C1 H4 O1 | 5e05755cc632ae24320df4abb87a324c | 14987 | 0 |
| C1 F3 O3 S1 1- | db7ef15f5ffc6dc283a42ec230113c73 | 10816 | -1 |
| N1 O3 1- | e68967e5fef3469d51de7c596f69f5d1 | 10096 | -1 |
| C1 H1 Cl3 | 49850386b8f6ea8e620755b3789b3c2d | 8953 | 0 |
| C4 H8 O1 | 80308d2c95b2814a59a763c56ce08576 | 8609 | 0 |
| C3 H7 N1 O1 | 3ea25a08243c42a0a19da4e2bcf10a27 | 8104 | 0 |
| C6 H6 | bbdf143f64d7c9cba2932fb985bfab55 | 7951 | 0 |
| C7 H8 | 50381c6ade55bbdc30b67f75a16339f9 | 7616 | 0 |
| C4 H10 O1 | b3c40d7be34a97dfc51939c446e13532 | 6229 | 0 |
| C3 H6 O1 | a8cc3e170fc2c659066c6c4a57a70748 | 5183 | 0 |
| C24 H20 B1 1- | 3683a785fbdf79707e93a5e009ee8c18 | 4608 | -1 |
| C16 H36 N1 1+ | 013c473064f36d0be2b2f3620531d885 | 4401 | 1 |
| C2 H6 O1 | a59b346cd7beb352d5b9bae0658151d7 | 4050 | 0 |
| C8 H20 N1 1+ | a5bdfc16c355f2842c06b82201034799 | 3797 | 1 |

**Text S3.** Iterative Complex Charge Workflow.

The charge assignment workflow determines component charges by enforcing electroneutrality at the level of individual crystallographic unit cells. Each unit cell is represented as a multiset of components (identified by graph hashes), and the total charge across all components is constrained to sum to zero. For any given unit cell, if all but one unique component has known charges (either from prior solved assignments or an initial manually assigned charge), the remaining unknown charge can be inferred by difference, accounting for component multiplicity within the cell.

These inferred charges are collected as independent observations ("votes") for each component across all unit cells in which it appears. Charge assignment proceeds iteratively: in each round, only unit cells containing exactly one unresolved component are used to generate new observations, and these observations are aggregated at the graph-hash level. After each round, candidate charges for each component are evaluated using a strict consensus criterion.

A component charge is finalized only when (i) at least three independent observations are available and (ii) the modal charge accounts for at least 67% of those observations. Components that do not meet these criteria remain unresolved but are tracked as solvable candidates.

Fractional values exactly at half-integers are discarded, and all accepted charges are restricted to the range −6 to +6.

Newly finalized charges are incorporated into the global set of known component charges and propagated to all unit cells in which those components appear. This reduces the number of unknowns in those cells and enables additional charge inferences in subsequent iterations. The procedure is repeated until no new charges can be finalized, at which point the system is considered converged. A final non-iterative pass is optionally performed to assign charges in any remaining unit cells where only a single unknown component remains, without applying consensus thresholds.

**Table S2.** Solvable charges per iteration.

| iteration | Solvable charges | Purity (median) |
|---|---|---|
| 1 | 428519 | 0.60 |
| 2 | 441714 | 0.65 |
| 3 | 450816 | 0.57 |
| 4 | 451509 | 0.60 |
| 5 | 451679 | 0.64 |
| 6 | 451706 | 0.67 |
| 7 | 451708 | 0.65 |
| 8 | 451709 | 0.65 |
| 9 | 451709 | 0.65 |

**Text S4.** Fractional Charge Handling in the Iterative Charge Assignment Workflow

To monitor how often fractional charge predictions required either rejection or rounding, we tracked the number of cases in each iteration that produced a half-integer value (i.e., an inferred charge ending in exactly 0.5) and the number that produced a non-integer value that was eligible for soft rounding. In iteration 1, 2,050 charge estimates fell exactly at a half-integer and were therefore rejected, while 309 fractional values were successfully rounded to the nearest allowed integer. In iteration 2, the counts increased to 2,774 half-integer rejections and 538 soft-rounded values. The pattern continued similarly in iteration 3 (2,923 half-integer cases, 569 soft-rounded values) and iteration 4 (2,943 and 572, respectively). By iteration 5, the totals stabilized at 2,948 half-integer cases and 578 soft-rounded values. These values remained unchanged for iterations 6 through 9.

For CSD refcode NIVSIY, the unit cell contains two copies of the complex of interest, along with 1 dichloromethane molecule and 1 tetrafluoroborate ion. Dichloromethane is neutral, so it does not contribute to the overall charge balance. The tetrafluoroborate ion carries a charge of –1. To balance this, each complex is assigned an effective charge of +0.5. Because this is a half-integer value, the algorithm classifies it as intrinsically ambiguous: it lies exactly between two plausible integer charge states and cannot be mapped cleanly onto a single, well-defined formal charge. In such cases, the charge inference procedure explicitly rejects the assignment rather than rounding up or down.

For CSD refcode AXOFOI, the unit cell contains three copies of the complex of interest and four triiodide counterions. Each triiodide carries a charge of –1, so the four counterions contribute a total charge of –4. To balance that, the three complexes together must contribute a total charge of +4. When this total is divided equally across the three identical complexes in the unit cell, each complex is assigned an effective charge of approximately +1.33. Because this inferred charge is neither an integer nor a half-integer, it falls into the category of fractional values that can be rounded safely. The algorithm therefore applies its soft-rounding rule and assigns the nearest allowed integer charge, which in this case is +1.

### a. BEJCUR02
### Iterative: 1  |  CSD: 1

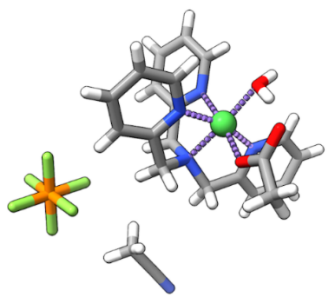

### b. SEFNEY
### Iterative: 0.5  |  CSD: 1

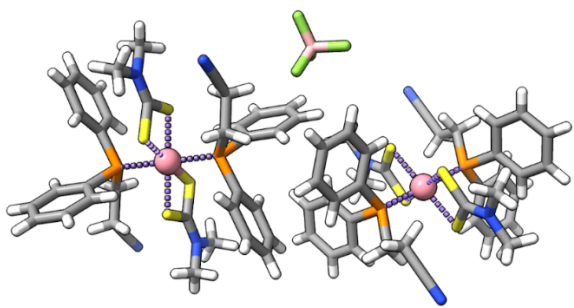

### c. MBZPNI
### Iterative: 0  |  CSD: 1

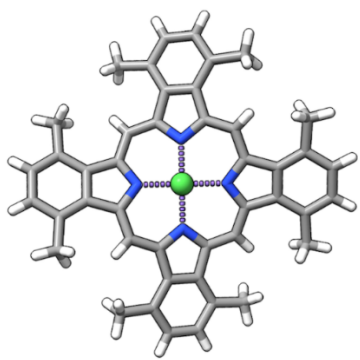

**Figure S2.** Iterative and CSD charge assignment comparison.
(a) Both methods agree on a +1 charge for the Ni (green) complex BEJCUR02. (b) For the Co (pink) complex SEFNEY, the CSD charge is correct, while the iterative method underestimates it because the deposited CIF contains two equivalent complexes, causing the mined charge to be halved. (c) For the Ni (green) porphyrin MBZPNI, the iterative method correctly assigns a neutral charge, whereas the CSD entry incorrectly reports +1.

**Text S5.** cell2mol-based oxidation state prediction workflow.

To supplement oxidation-state assignments beyond those obtained through text parsing of the CSD metadata, we employed cell2mol[4] on 43,625 complexes. For each crystallographic entry in the curated CSD subset, the raw CIF text was written temporarily to disk and processed through cell2mol using a wrapper routine with a 120-second timeout. The resulting .gmol output was examined to extract molecular components identified as complexes. For each detected complex, the oxidation state of the metal center was collected together with the corresponding CSD refcode and atom label, generating a structured table of inferred metal oxidation states. Entries that failed to parse successfully or lacked a recognizable transition-metal complex were logged separately. Across the processed entries, cell2mol provided 16,433 additional unique oxidation-state assignments, which were merged with the oxidation states extracted by the text-based parser described in the main text.

**Text S6.** Details for isolating ligands from complexes.

To isolate individual ligands from each mononuclear transition-metal complex, we decomposed the molecular graph into non-metal connected components using a NetworkX[5]-based procedure. Each structure was first converted into a graph representation where atoms correspond to nodes and covalent bonds to edges. Metal atoms were identified using the molSimplify[2,3] built-in metal-detection routine and treated as coordination centers rather than components of ligand fragments.

The ligand breakdown proceeded in three steps:

1. Identification of metal–ligand connecting atoms.
   For each metal center, we enumerated all atoms directly bonded to the metal. These "connecting atoms" define the attachment points between each ligand fragment and the metal.

2. Removal of metal nodes and graph partitioning.
   The metal atoms were removed from the molecular graph, and the remaining graph was partitioned into its connected components. Because the dataset contains only mononuclear complexes, each connected component corresponds to a single terminal ligand fragment.

3. Construction of ligand submolecules and metadata.
   For each identified connected component, a sub-molecule was constructed containing only the ligand atoms. We then recorded the indices of coordinating atoms within the ligand (relative to the subgraph), and computed a Weisfeiler–Lehman (WL) hash[6] for the ligand using atom symbols as node attributes. This hash provides a topology-based identifier that clusters chemically identical ligands across complexes, independent of geometry or protonation state.

A helper routine validated that no connected component contained multiple metals, confirming the absence of bridging ligand motifs in the mononuclear dataset. The output of the ligand-breakdown procedure is a list of ligand submolecules paired with metadata including (i) its coordinating atom indices, and (ii) its WL graph hash. This standardized representation is used throughout the subsequent ligand-charge inference workflow.

**Text S7.** Weighting scheme used for ligand charge assignment.

Candidate ligand charges were not treated equally across all observations. Instead, each charge assignment was recorded together with a reliability weight designed to favor chemically and crystallographically cleaner cases. In practice, the weight for a given observation was based on three factors: the crystallographic quality of the parent structure, the number of distinct ligand types present in the complex, and whether the charge deduced for the ligand was an integer.

For a given complex, the initial weight was defined as

$$w_0 = \frac{1}{N_{\text{types}}} \cdot \frac{1}{1 + (R/100)\alpha}$$

where $N_{\text{types}}$ is the number of unique ligands present in the ligand breakdown for that complex, $R$ is the reported CSD $R$-factor in percent, and $\alpha$ is a scaling factor (set to 10 in this work). This form gives greater influence on assignments from simpler heteroleptic environments and from better-resolved crystal structures. If no $R$-factor was available, the structure was assigned a default value of 100, which strongly downweighted that observation.

Ligand charges were inferred by charge balance. For each complex, the ligand contribution was obtained from the total formula charge after subtracting the metal oxidation state and the charges of ligands that had already been assigned in earlier steps of the iterative procedure. When the resulting charge for the unresolved ligand was an exact integer, the full weight $w_0$ was retained. When the inferred value was non-integer but not exactly half-integer, the charge was rounded to the nearest integer, and its contribution was penalized by multiplying the weight by 0.5. In contrast, assignments that gave exact half-integer values were discarded by setting their weight to zero, since these cases were taken to indicate an internally inconsistent or ambiguous charge-balance solution.

Each ligand graph hash could therefore accumulate multiple charge observations of the form (assigned charge, source refcode, weight, unrounded charge). The final active charge for a ligand was then chosen as the charge state with the largest summed weight across all observations, rather than by simple majority count. This weighted-voting procedure was intended to prioritize assignments supported by high-quality structures and less ambiguous heteroleptic environments while still allowing weaker evidence to contribute when consistent with the overall dataset.

**Table S3.** Solved ligand charges per iteration.

| iteration | Solved charges |
|---|---|
| **1 (Homoleptic)** | 16686 |
| **2** | 66162 |
| **3** | 66791 |
| **4** | 66809 |
| **5** | 66810 |

**Text S8.** Coverage of total, and solvable ligand space.

To quantify the coverage of our ligand-charge inference workflow, we used the ligand-breakdown procedure (Text S4) to enumerate all ligand fragments across every mononuclear TMC in the curated dataset. This produced 94,581 unique WL-hashed ligand graphs, representing the total ligand chemical space, regardless of whether the parent complex had a resolvable charge or oxidation state.

Because ligand-charge inference requires both a known complex charge and a defined metal oxidation state, we next restricted the dataset to the subset of complexes that passed all complex charge preprocessing steps. Applying the same ligand-breakdown procedure to this filtered set yielded 68,380 unique ligands, corresponding to the solvable ligand space—i.e., ligands for which charge determination was in principle possible. Our iterative workflow successfully assigned charges to 66,810 ligands, representing 97% coverage of the solvable ligand set.

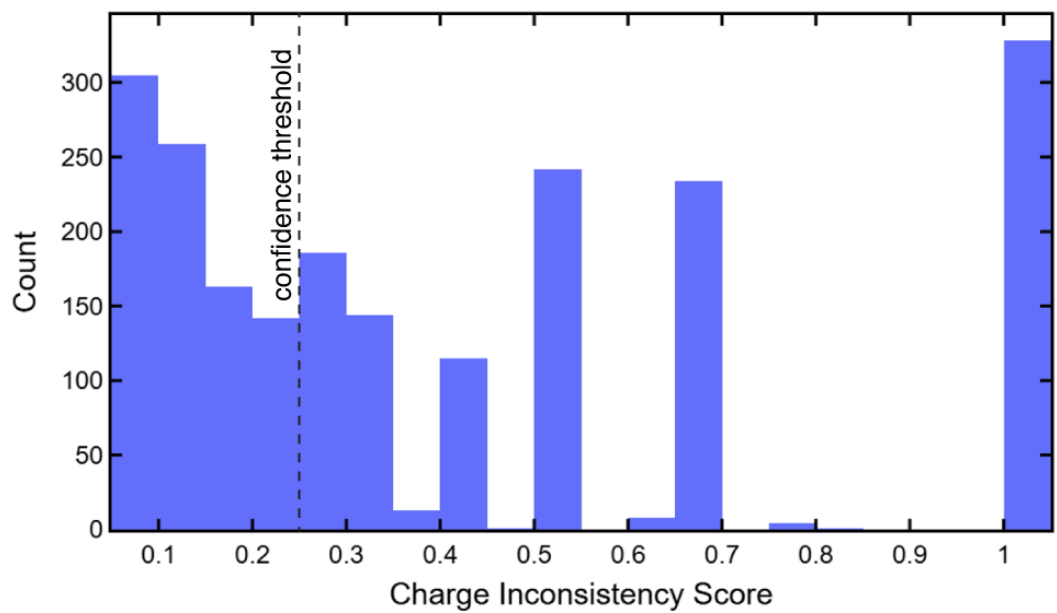

**Figure S3.** Charge inconsistency score amongst mined ligands. All ligands with no charge inconsistency were removed. Ligands in with low charge inconsistency scores are ligands that commonly appear across the CSD, have a handful of bad entries across their appearances. The weighting scheme introduced primarily helped with determine charge among ligands that had high inconsistency (> 0.25).

**Text S9.** Obtaining Titles and Abstracts for CSD entries.

Bibliographic records were obtained by querying the Web of Science API[7] using DOIs associated with CSD entries. Of 94,871 DOI queries, abstracts and associated metadata were successfully retrieved for 81,948 records (86.4%), while 12,923 records (13.6%) lacked abstract-level metadata. Missing records were concentrated primarily from a few particular sources. Inorganic Chemistry (2,088 records), Journal of the American Chemical Society (1,635 records), Dalton Transactions (1,123 records), Angewandte Chemie International Edition (951 records), and Journal of Organometallic Chemistry (942 records) together account for 52.1% of all missing entries. We anticipate these articles are missing abstracts either because the letters published in these journals did not have abstracts or because the text was not suitably digitized. Other missing records were not contained in the Web of Science.

**Text S10.** Topic Modeling and Classification.

We adopted a BERTopic-based[8] workflow as in prior work[9], with select modifications. Using this approach, we curate application-linked ligand subsets for the following application areas: reactivity and catalysis (react), biological (bio), spin-state and magnetism (magnet), redox chemistry (redox), and photophysical (photo) chemistry. Topic inference was guided rather than fully unguided. Instead of discovering topics and assigning application labels post hoc, we seeded the class-based Term Frequency-Inverse Document Frequency (c-TF-IDF) representation with curated keyword sets corresponding to reactivity, biological, photophysical, redox, and spin state and magnetism applications[8]. This seeding biases topic formation toward chemically meaningful functional domains while preserving unsupervised clustering at the document level. Topic-level assignments were complemented by document-level semantic similarity scoring, allowing multi-label classification when multiple application contexts were strongly represented. To reduce spurious multi-label assignments, we applied an additional evidence-based post-processing step specifically when theme labels competed within the same document. We extended the workflow to include time-resolved analysis of both topics and derived ligand usage by incorporating publication year metadata and aggregating results into fixed temporal bins. This enables direct interrogation of how application emphasis and ligand reuse evolve over time, an aspect not explored in prior work.

**Text S11.** Ligand Topic Purity Scheme.

To quantify how selectively a ligand is associated with a functional domain, we define a purity score based on how that ligand's literature-linked occurrences are distributed across five primary application areas (reactivity, redox chemistry, biological, photophysical, and spin state/magnetism). For each ligand, we consider the set of associated bibliographic records (each providing a single application label from our topic-classification workflow) and count how many times the ligand appears in each application area, treating each ligand–record co-occurrence as one appearance. The purity of a ligand with respect to a given application is then computed as the fraction of that ligand's total appearances across the five primary applications that fall within the area of interest (i.e., the application-specific count divided by the sum of that ligand's counts over all five applications). Interpreting this metric is straightforward: purity values near 1 indicate ligands that are specialized to a single application domain, whereas values near 0 reflect broad reuse across multiple domains. The governing equation is:

$$p_{l,a} = \frac{n_{l,a}}{\sum_{a' \epsilon A} n_{l,a'}}$$ (S1)

$$A = \{\text{react, bio, magnet, redox, photo}\}$$ (S2)

*A* is the complete set of application areas, *n*, is the number of events a given ligand (*l*) is classified into application area (*a*), and this is over the summation of every event that a given ligand has among every application.